\documentclass[superscriptaddress, aps, pre, reprint, floatfix]{revtex4-2}
\usepackage{graphicx,hyperref,color,upgreek}
\usepackage{amsmath}
\usepackage{latexsym}
\usepackage{float}
\usepackage{amssymb}
\usepackage{multirow}
\usepackage{graphicx}
\usepackage{textcomp}
\usepackage{hyperref}
\usepackage{array}
\usepackage{xcolor}
\usepackage{hyperref}
\usepackage{xkeyval,xcolor}
\usepackage{xcite}
\usepackage[caption=false]{subfig}
\usepackage{dcolumn}
\usepackage{bm}
\usepackage{xr}
\usepackage{tikz}
\makeatletter
\newcommand*{\addFileDependency}[1]{
  \typeout{(#1)}
  \@addtofilelist{#1}
\IfFileExists{#1}{}{\typeout{No file #1.}}
}
\newcommand{\subfigimg}[3][,]{%
  \setbox1=\hbox{\includegraphics[#1]{#3}}
  \leavevmode\rlap{\usebox1}
  \rlap{\hspace*{40pt}\raisebox{\dimexpr\ht1-2\baselineskip}{#2}}
  \phantom{\usebox1}
}
\makeatother

\newcommand*{\myexternaldocument}[1]{%
    \externaldocument{#1}%
    \addFileDependency{#1.tex}%
    \addFileDependency{#1.aux}%
}

\myexternaldocument{abp_fieldSI}

\begin{document}


\title{Phase Behavior and Dynamics of Active Brownian Particles in an Alignment Field}
\author{Sameh Othman}
\email{s.othman@fz-juelich.de}
\affiliation{
Theoretical Physics of Living Matter, Institute of Biological Information Processing and Institute for Advanced Simulation, Forschungszentrum J\"{u}lich,  52425 J\"{u}lich, Germany
}
\author{Jiarul Midya}
\email{jmidya@iitbbs.ac.in}
\affiliation{
Theoretical Physics of Living Matter, Institute of Biological Information Processing and Institute for Advanced Simulation, Forschungszentrum J\"{u}lich,  52425 J\"{u}lich, Germany
}
\affiliation{
School of Basic Sciences, Indian Institute of Technology, Bhubaneswar, 752050, India
}
\author{Thorsten Auth}
\email{t.auth@fz-juelich.de}
\affiliation{
Theoretical Physics of Living Matter, Institute of Biological Information Processing and Institute for Advanced Simulation, Forschungszentrum J\"{u}lich,  52425 J\"{u}lich, Germany
}
\author{Gerhard Gompper}
\email{g.gompper@fz-juelich.de}
\affiliation{
Theoretical Physics of Living Matter, Institute of Biological Information Processing and Institute for Advanced Simulation, Forschungszentrum J\"{u}lich,  52425 J\"{u}lich, Germany
}

\date{\today}

\begin{abstract}
Self-propelled particles that are subject to noise are a well-established generic model system for active matter. A homogeneous alignment field can be used to orient the direction of the self-propulsion velocity and to model systems like phoretic Janus particles with a magnetic dipole moment or magnetotactic bacteria in an external magnetic field. Computer simulations are used to predict the phase behavior and dynamics of self-propelled Brownian particles in a homogeneous alignment field in two dimensions. Phase boundaries of the gas-liquid coexistence region are calculated for various P\'eclet numbers, particle densities, and alignment field strengths. Critical points and exponents are calculated and, in agreement with previous simulations, do not seem to belong to the universality class of the 2D Ising model. Finally, the dynamics of spinodal decomposition for quenching the system from the one-phase to the two-phase coexistence region by increasing P\'eclet number is characterized. Our results may help to identify parameters for optimal transport of active matter in complex environments. 
\end{abstract}

\maketitle


\section{\label{sec:introduction}Introduction}

\begin{figure}
    \centering
    \includegraphics[width=\columnwidth]{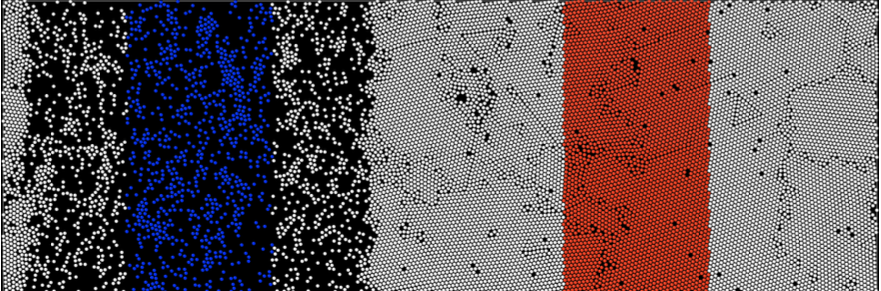}
    \caption{Simulation snapshot of a system that contains $12,576$ particles in a box with dimensions of $L_x=80\sigma$, $L_y=240\sigma$ at $\rm Pe=144$ with an alignment field of strength $\tilde{B}=2$ in $y$-direction. Two stacked square boxes with side lengths $\ell_{\rm s}=40\sigma$, placed at the centers-of-mass of the liquid (red) and the gas phase (blue), are used to measure the particle densities.}
    \label{fig:illustration}
\end{figure}

Active matter is abundant in life and ranges from the cytoskeleton \cite{schaller_polar_2010}, tissues \cite{xi_material_2019}, and biofilms \cite{jeckel_shared_2022} on the microscale to fish schools \cite{Parrish_SelfOrganized_2002}, animal herds \cite{garcimartin_flow_2015}, and pedestrian crowds \cite{chraibi_generalized_2010} on the macroscale. In synthetic and engineered systems, phoretic Janus particles are a versatile model system on the microscale \cite{fadda_interplay_2023,roca-bonet_self-phoretic_2022,vutukuri_active_2020,volpe_microswimmers_2011}, and small robots and vibrated granular matter on the macroscale \cite{giomi_swarming_2013,deseigne_collective_2010}. The systems consist of individual self-propelled agents, thus are intrinsically out-of-equilibrium and show complex emergent collective behavior \cite{elgeti_physics_2015,bechinger_active_2016}. 
However, there is no general concept for predicting the phase behavior analogous to the minimization of the free energy at thermal equilibrium.
Only in very few cases, such as for spherical and rod-like active Brownian particles and filament-motor mixtures, analytical expressions have been proposed to predict phase behavior \cite{stenhammar_continuum_2013,baskaran_hydrodynamics_2008,doostmohammadi_stabilization_2016}. Therefore, high-performance computing is often the method of choice to characterize the structure and dynamics of non-equilibrium systems.

A well-established generic model system for dry active matter is active Brownian particles (ABPs), whose motion, in addition to their self-propulsion velocity, is subject to thermal noise.
Intriguingly, ABPs exhibit motility-induced phase separation (MIPS) at high densities, blocking each other's motion and leading to cluster formation. 
Analogous to the vapor-liquid transition in passive systems, high-density liquid clusters of ABPs coexist with a low-density active gas phase, however, without any attraction. 
Cooperative motion has been detected in three-dimensional dense suspensions of active Brownian particles despite the lack of an alignment mechanism. It has been hypothesized that this collective swirling motion is driven by an interface-sorting process \cite{wysocki_cooperative_2014}. 

Critical points characterized by power laws with universal exponents are a prevalent aspect of second-order phase transitions, such as the critical point of the gas-liquid coexistence. 
Previous simulation studies of two-dimensional (2D) ABP systems have accurately mapped out the binodals and estimated the location of the critical point \cite{dittrich_critical_2021}. It has been shown that the associated critical exponents are different from the standard 2D Ising universality class. 
In contrast, the dynamical critical exponent, related to the relaxation dynamics of the system after a quench from a homogeneous state below the critical point to the two-phase region, is consistent with the 2D Ising universality class \cite{dittrich_growth_2023}. The controversy regarding the universality class remains alive and demands further investigation.

An interesting system is active Brownian particles in an alignment field for the direction of their self-propulsion, see Fig.~\ref{fig:illustration}. An experimental realisation of such a system are magnetic particles with a dipole moment that can couple to an external magnetic field, which is routinely used for magnetic microrheology \cite{Hess_Scaledependent_2020} and which can be employed to generate anisotropic elastic materials \cite{Braunmiller_PreProgrammed_2022}. 
For low densities and sufficiently weak magnetic dipole moments, the mutual interaction between magnetic colloids can be neglected, whereas for high densities and strong dipole moments a magnetic alignment of the dipoles has to be taken into account \cite{zinn_dynamics_2023}. However, an alignment of the direction of the propulsion velocity does not necessarily have to be due to magnetic-dipole interactions or even follow any physical interaction rules. For example, also an alignment rule inspired by the ferronematic four-state Potts model \cite{chatterjee2020flocking}, excluded-volume interactions of particles with elongated shapes \cite{abkenar_collective_2013,peruani_nonequilibrium_2006}, 
and visual perception may lead to a velocity alignment \cite{negi_collective_2024,barberis_large-scale_2016}. The most prominent example is the Vicsek model, where alignment interactions between self-propelled agents orient their direction of motion with respect to the orientations of their neighbours \cite{vicsek_novel_1995}. 

In this work, we use Brownian Dynamics simulations to study the critical behavior of two-dimensional ABP systems subject to a homogeneous external alignment field that couples to the polar direction of the ABP self-propulsion velocities. 
The phase-separated systems show characteristic stripe patterns that are oriented parallel to the direction of the field. In addition to steady states, we simulate and analyze the domain coarsening dynamics after quenches from the one-phase to the two-phase region. 
Here, we distinguish between the directions parallel and perpendicular to the field. 
For the striped structures, we find a much faster growth of cluster size in the direction parallel to the alignment field compared with the direction perpendicular to the field.

The remainder of the manuscript is organized as follows. In section \ref{sec:model}, we introduce the system and the simulation technique. In section \ref{sec:phases}, we predict the two-phase fluid-gas coexistence region, the critical points, and the critical exponents for various alignment field strengths. In section \ref{sec:dynamics}, we discuss spinodal decomposition for a quench from points in the phase space with P\'eclet numbers below the critical point into the two-phase coexistence region.
Finally, in section \ref{sec:conclusions} we summarize our results and provide an outlook.

\section{\label{sec:model}Model and Methods}

We use Brownian Dynamics simulations to simulate the motion of ABPs in two dimensions in an external alignment field $\mathbf{\tilde{B}}$, which is governed by
\begin{align}
    \mathbf{\dot r} & = \sqrt{2\rm D_{\rm T}}\boldsymbol{\xi}_{\rm T} + \gamma_{\rm T}^{-1}\mathbf{F} + v_0\mathbf{\hat{e}} \\
    \mathbf{\dot{\hat{e}}} & = \sqrt{2\rm D_{\rm R}}\mathbf{\hat{e}} \times \boldsymbol{\xi}_{\rm R} + \gamma_{\rm R}^{-1} \mu \mathbf{\hat{e}} \times (\mathbf{\hat{e}} \times \mathbf{\tilde{B}})\, ,
\end{align}
where $\mathbf{\dot r}$ is the translational and $\mathbf{\dot{\hat{e}}}$ is the angular velocity and $\gamma_{\rm T}$ and $\gamma_{\rm R}$ are the translational and rotational friction coefficients, respectively. 
The Gaussian-distributed random noises with unit variance, $\rm \boldsymbol{\xi}_{\rm T}$ and $\rm \boldsymbol{\xi}_{\rm R}$, and are vectors in the 2D plane in that the particles move. Each particle possesses a self-propulsion velocity $v_0 \mathbf{\hat{e}}_i$, which is constant in magnitude and  parallel to the dipole moment $\boldsymbol{\mu} = \mu \mathbf{\hat{e}}_i$ of the particle that couples to an external alignment field $\mathbf{B}$. 

The particle-particle interaction is taken into account by the force $\mathbf{F}$, which is derived from the purely repulsive Weeks-Chandler-Andersen (WCA) potential
\begin{equation}
        \rm U_{\rm WCA}(r)=\left\{
        \begin{matrix}
        \,\, 4\epsilon[(\frac{\sigma}{r})^{12}-(\frac{\sigma}{r})^{6}+1/4]\,\,\, & r<2^{1/6}\sigma\\
        0\,\,\, & r\geq 2^{1/6}\sigma \, ,
        \end{matrix}
        \right.
\end{equation}
where $r$ is the distance between two particles, and $\epsilon$ characterizes the height and $\sigma$ the width of the potential. The potential vanishes at $r_{\rm min} = 2^{1/6}\sigma$, where we truncate the potential. 

We use the P\'eclet number ${\rm Pe}=3v_0/(d_{\rm BH} D_{\rm R})=v_0 d_{\rm BH}/D_{\rm T}$ to characterize the self-propulsion. Here, the translational diffusion coefficient is $D_{\rm T}=k_B{\rm T}/\gamma_{\rm T}$ with the Boltzmann constant $k_B$ and the temperature $\rm T$. 
The relation between the rotational and translational diffusion coefficients is $D_{\rm R}=3 D_{\rm T}/d_{\rm BH}^{2}$, which applies to spherical colloids in a Newtonian fluid in $3$D \footnote{For spherical particles and 3D hydrodynamics, the translational diffusion coefficient is $D_{\rm T}=k_B\rm T/\gamma_{\rm T}$ and the translational drag coefficient $\gamma_{\rm T} = 6\pi\eta R$, where $k_{\rm B}$ is the Boltzmann constant, $T$ the absolute temperature, $\eta$ the fluid viscosity, and $\rm R$ the particle radius. 
The rotational motion of the particle is described by $\boldsymbol{\dot{\varphi}}=\sqrt{2\rm D_R}\boldsymbol{\xi}_R$ with the rotational diffusion coefficient $D_{\rm R}=k_{\rm B}T/\gamma_R$ and the rotational drag coefficient $\gamma_{\rm R} = 8\pi\eta R^3$.}. 
Here, the effective particle radius at thermal equilibrium is defined as the Barker-Henderson radius $d_{\rm BH}=\int_{0}^{r_{\rm min}} dr\, [1-\exp(-U_{\rm WCA}(r))]$. 
By setting $\epsilon=100 \, k_B\rm T$, we ensure the particles behave similar to hard spheres with an effective radius $d_{\rm BH}=1.10688\sigma$. 
In the following, we use the characteristic timescale $\tau_R=1/D_{\rm R}$ for the rotation of the particle dipole moments as a time unit.

For all simulations, we initialize the systems by placing $N$ particles randomly in rectangular or square simulation boxes with periodic boundary conditions at various particle packing fractions $\phi=\pi d_{BH}^2 \rho/4$, where $\rho$ is the number density of the particles. 
We use LAMMPS to perform the simulations, see appendix \ref{sec:LAMMPS} for the values of the simulation parameters and further details of the simulations.

\section{\label{sec:phases}Phase diagrams and critical behavior}

\begin{figure}
  \centering
  \begin{tabular}{@{}p{0.9\columnwidth}@{\quad}p{0.9\columnwidth}@{}}
    \subfigimg[width=\linewidth]{(a)}{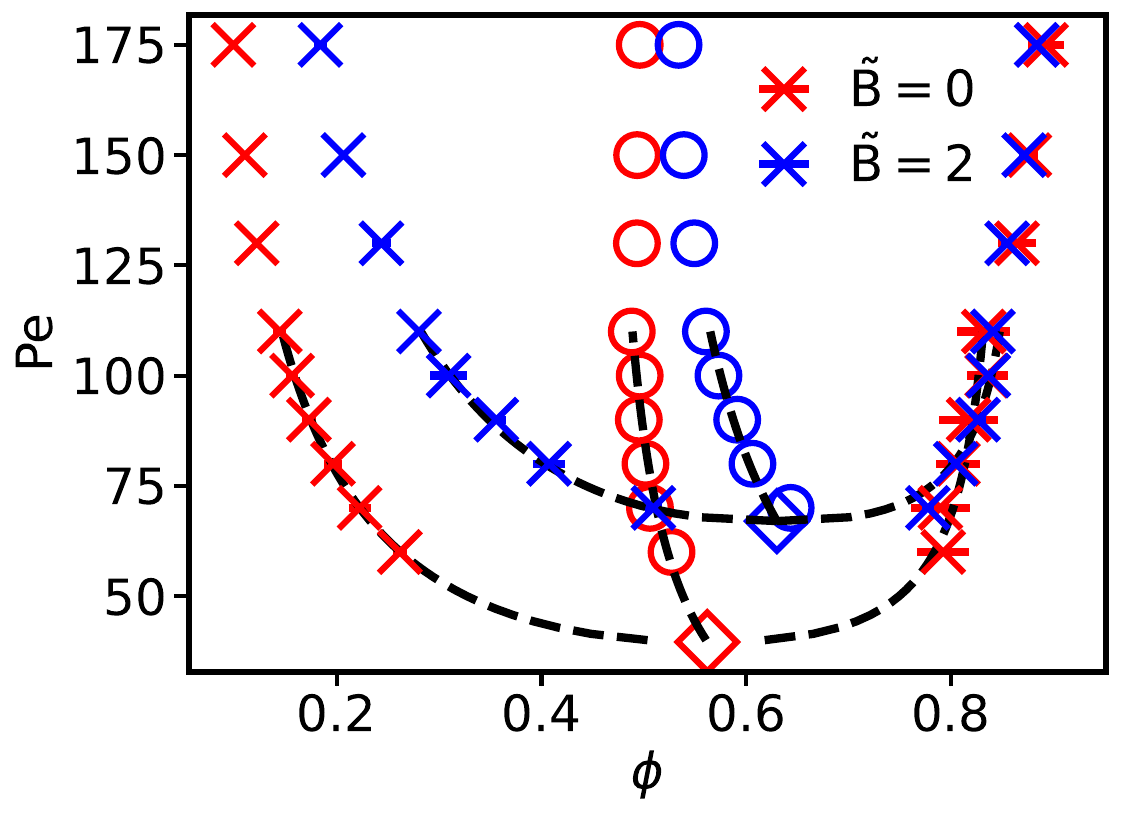} \quad \\
    \subfigimg[width=\linewidth]{(b)}{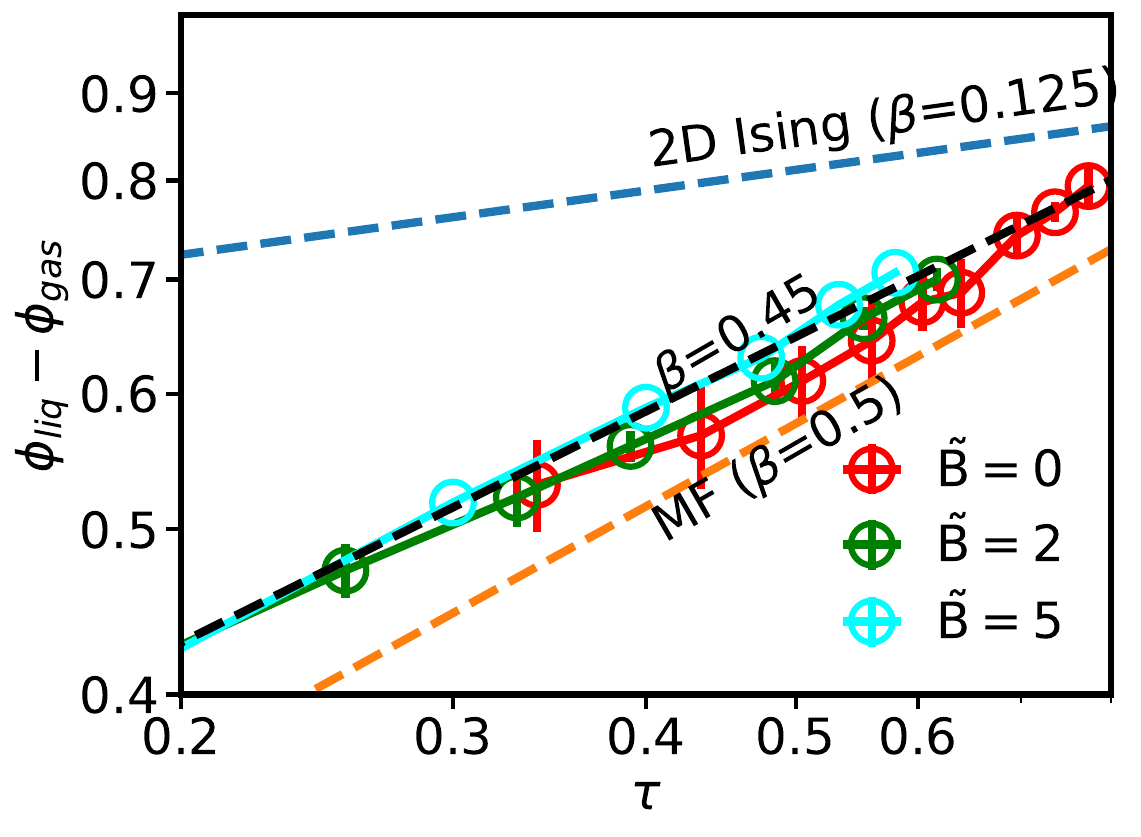} \quad \\
  \end{tabular}
    \caption{Critical points of active-Brownian-particle systems. (a) Boundaries for the gas-liquid coexistence region for the alignment field strengths $\tilde{B}=2$ (blue) and $\tilde{B}=0$ (red). Crosses mark the phase boundaries, diamonds the critical points, and circles the values of the rectilinear diameter $\phi_{\rm d}$, compare Eq.~(\ref{eq:phid}). (b) Order parameter 
    $\phi_{\rm liq}-\phi_{\rm gas}$ vs. the distance to the critical point $\tau$ for alignment fields $\tilde{B}=0$, $2$, and $5$ and measured power-law exponents $\beta$. The exponents for $2$D Ising and mean-field (MF) systems are shown for comparison. 
    }
  \label{fig:phase_diagram}
\end{figure}
\begin{figure}
  \centering
  \begin{tabular}{@{}p{0.9\columnwidth}@{\quad}p{0.9\columnwidth}@{}}

    \subfigimg[width=\linewidth]{(a)}{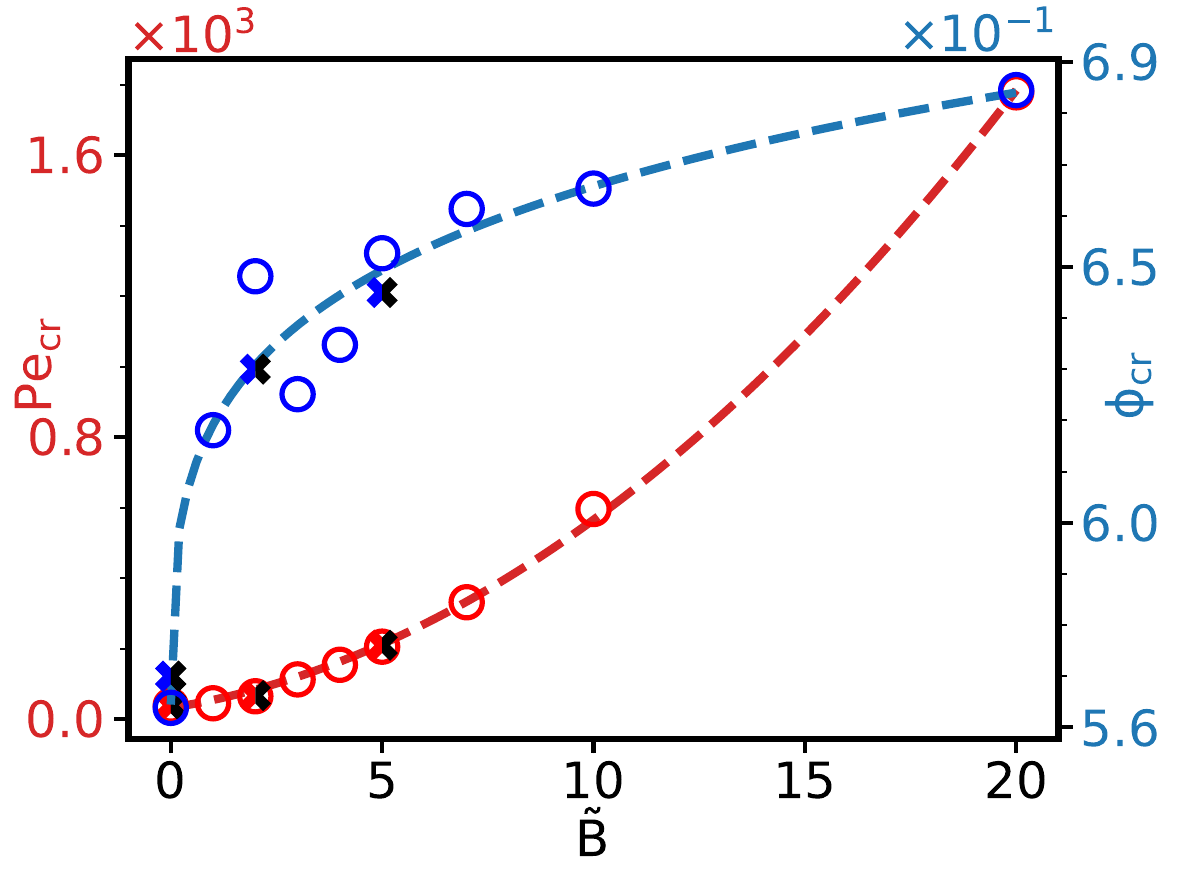} \quad \\
    \subfigimg[width=\linewidth]{(b)}{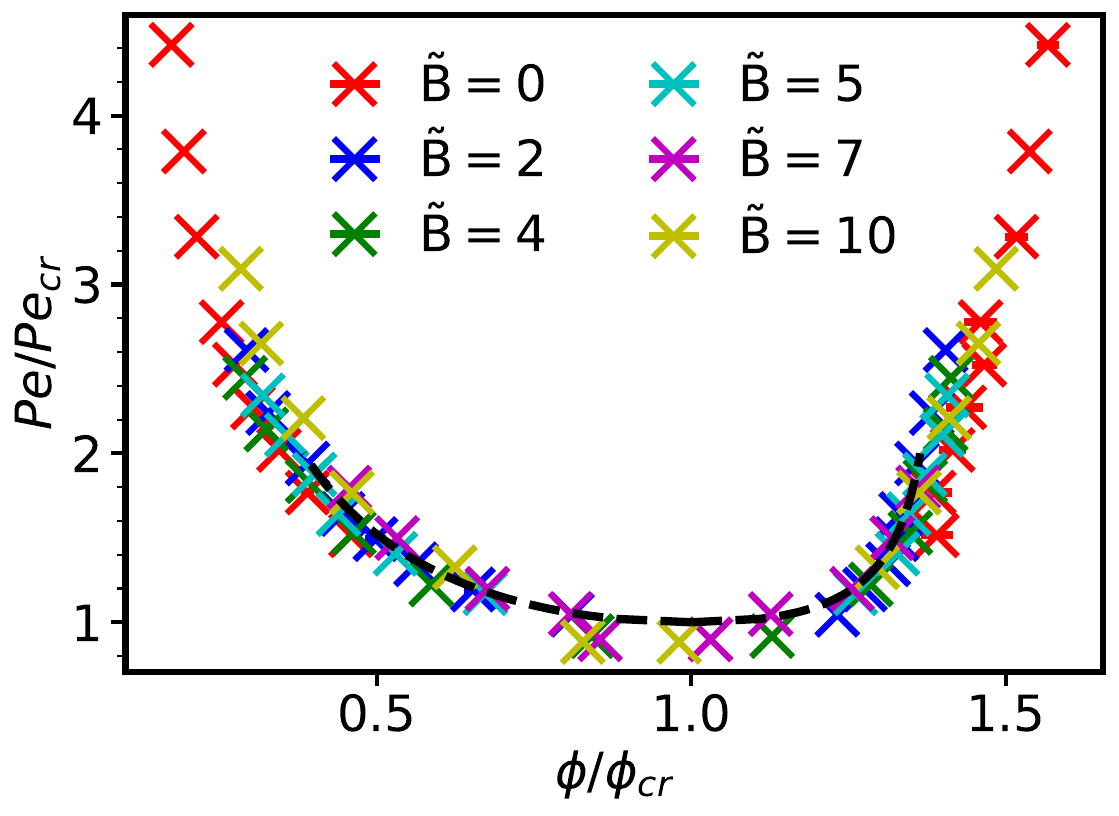} 
  \end{tabular}
  \caption{Critical point positions and boundaries of the two-phase coexistence region for various alignment field strengths. (a) Coordinates of the critical points as function of the alignment field strength $\tilde{B}$, determined using cumulants (crosses) or by assuming $\beta=0.45$ and simultaneously fitting the data using Eqs.~(\ref{eq:philmphig}) and (\ref{eq:phid}). 
  (b) Normalized coexisting phase diagrams for various values of $\tilde{B}$ and fit for $\tilde{B}=0$.}
  \label{fig:phase_critical}
\end{figure}

Above a critical P\'eclet number $\rm Pe_{cr}$ and for intermediate packing fractions $\phi$, we observe the coexistence of a high-density liquid phase and a low-density gas phase. We calculate the packing fractions of both phases from one simulation at values of  $\rm Pe$ and $\phi$ in the two-phase region, see Figs.~\ref{fig:illustration} and \ref{fig:phase_diagram}(a). Without an alignment field ($\tilde{B}=0$), the coexisting packing fractions of the gas and the liquid phase agree well with those reported previously \cite{siebert_critical_2018}. 
The difference between the liquid and gas packing fractions increases with increasing dimensionless "distance" $\tau = (\rm Pe_{\rm cr}^{-1} - Pe^{-1})/\rm Pe_{\rm cr}^{-1}$ from the critical point, 
\begin{equation}
    \phi_{\rm liq}-\phi_{\rm gas} = c_1\tau^\beta
    \label{eq:philmphig}
\end{equation}
with the critical exponent $\beta$. The rectilinear diameter
\begin{equation}
    \phi_{\rm d} = \frac{\phi_{\rm liq}+\phi_{\rm gas}}{2} \label{eq:phid}
    = \phi_{\rm cr} + c_2 \tau + O \left( \tau^2 \right) 
\end{equation}
is the arithmetic mean of $\phi_{\rm gas}$ and $\phi_{\rm liq}$ and
ends at the critical point at $\phi_{\rm cr}$ and $\rm Pe_{\rm cr}$.
In the presence of an alignment field, the area of the two-phase region shrinks compared to the system without field, see Fig.~\ref{fig:phase_diagram}(a). 
Whereas at high values of $\rm Pe$ the boundary of the two-phase region at high packing fractions is almost unchanged compared with $\tilde{B}=0$, the boundary at low packing fractions significantly shifts to higher $\phi$. 
Therefore, for fixed $\rm Pe$, the rectilinear diameter $\phi_{\rm d, \tilde{B}=2}$ increases with increasing field strength. 

\begin{figure}
  \centering
  \subfloat{\includegraphics[width=.4\columnwidth]{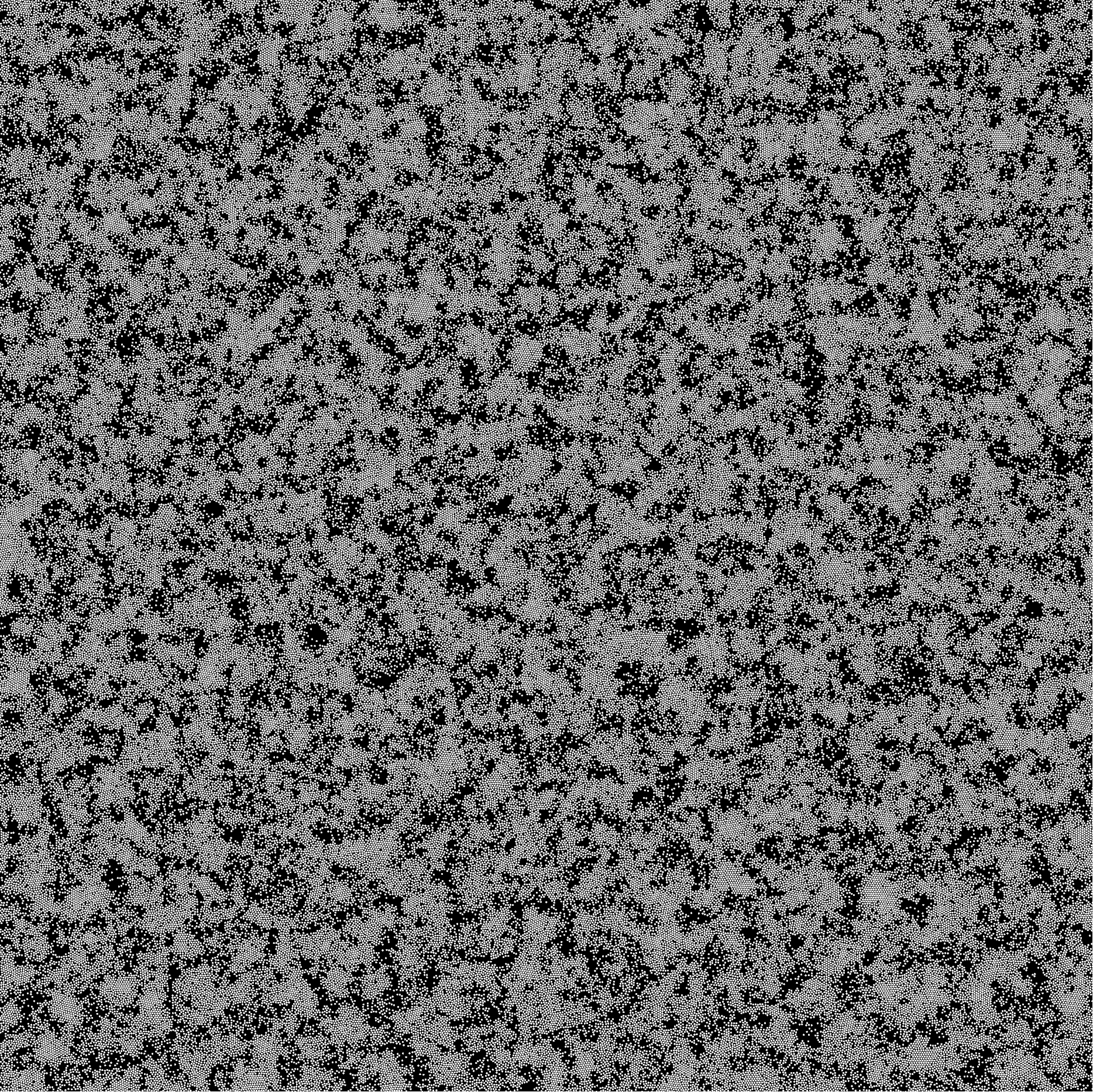}}\quad
  \subfloat{\includegraphics[width=.4\columnwidth]{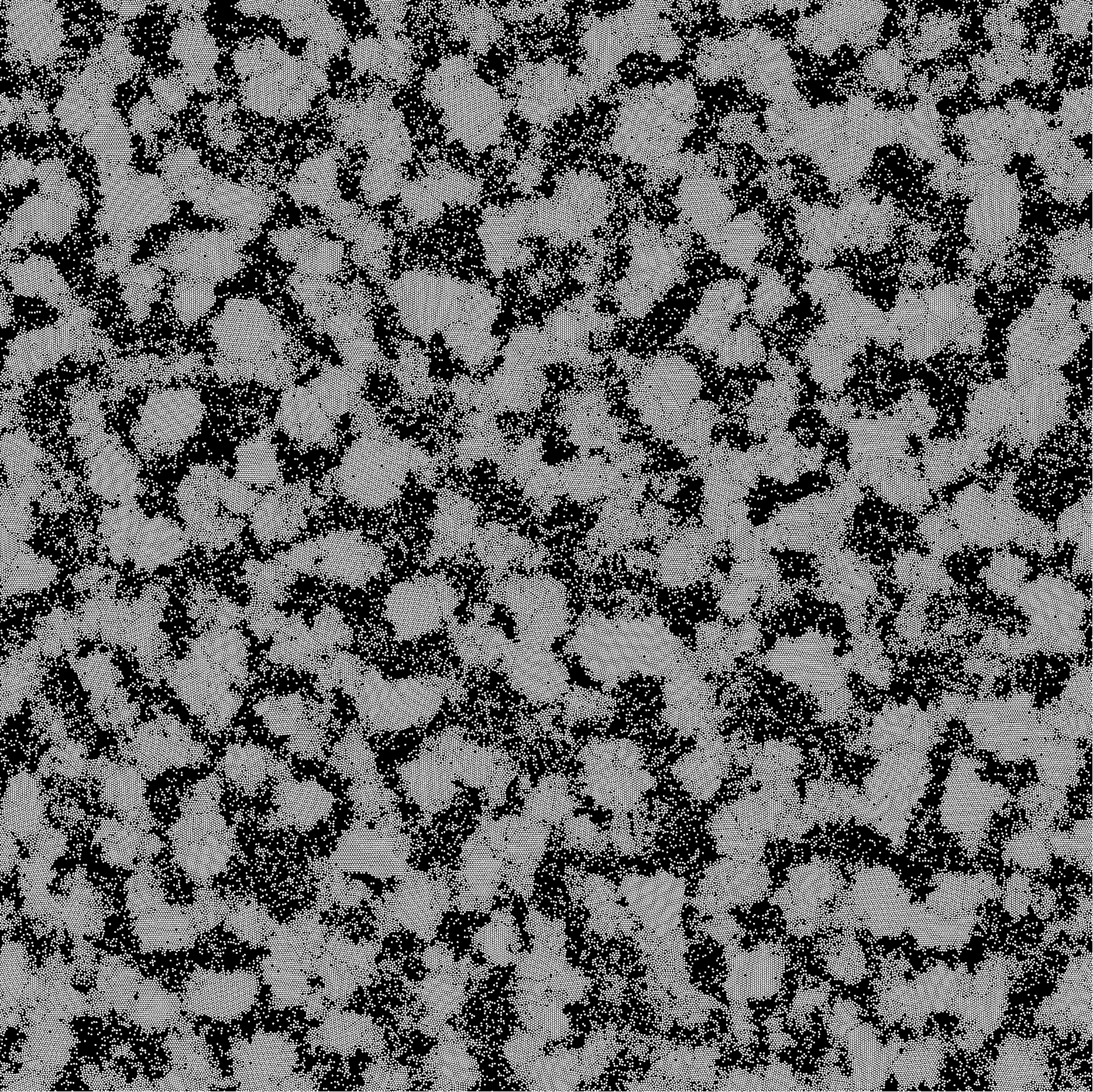}}\\
  \subfloat{\includegraphics[width=.4\columnwidth]{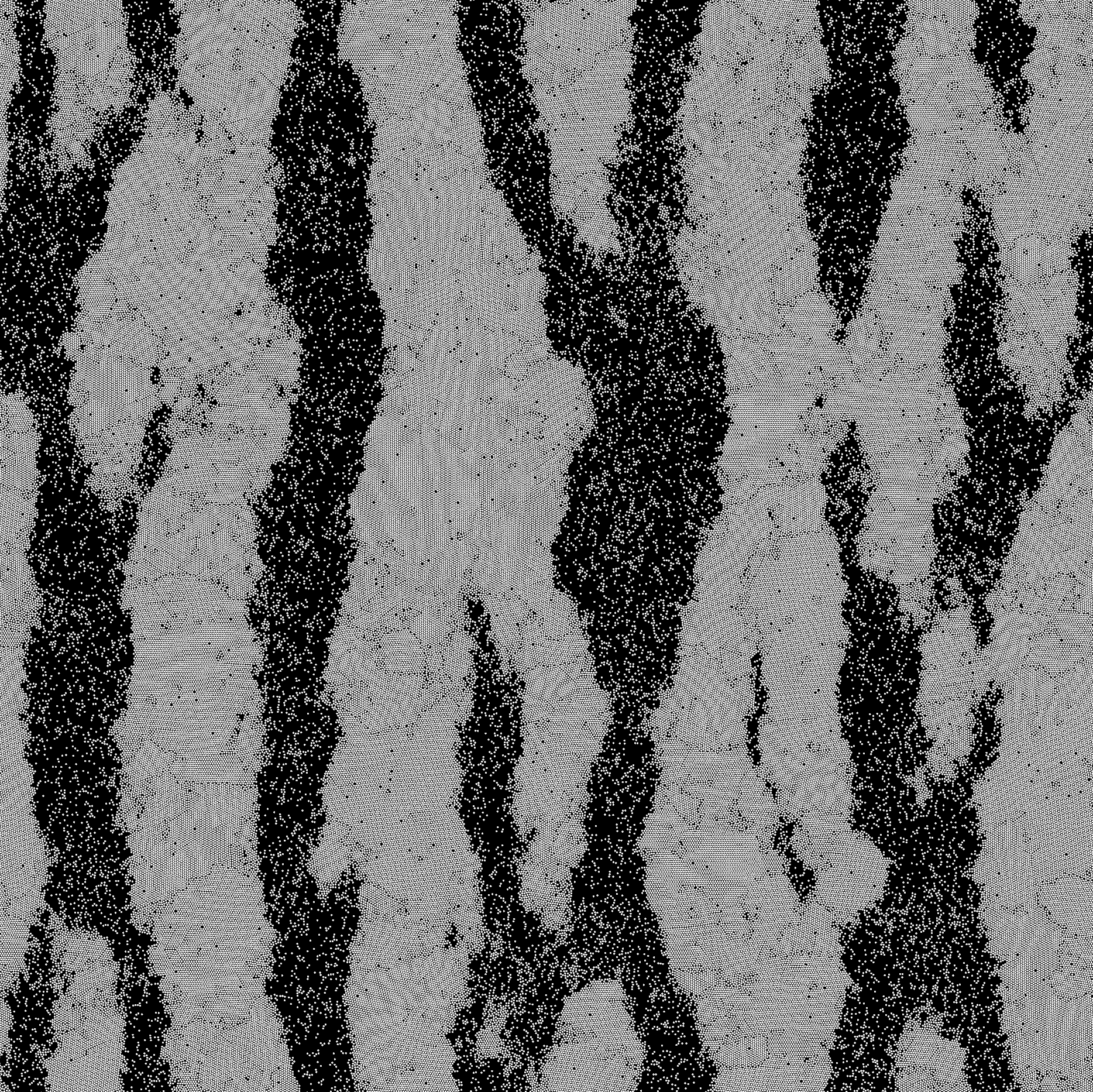}}\quad
  \subfloat{\includegraphics[width=.4\columnwidth]{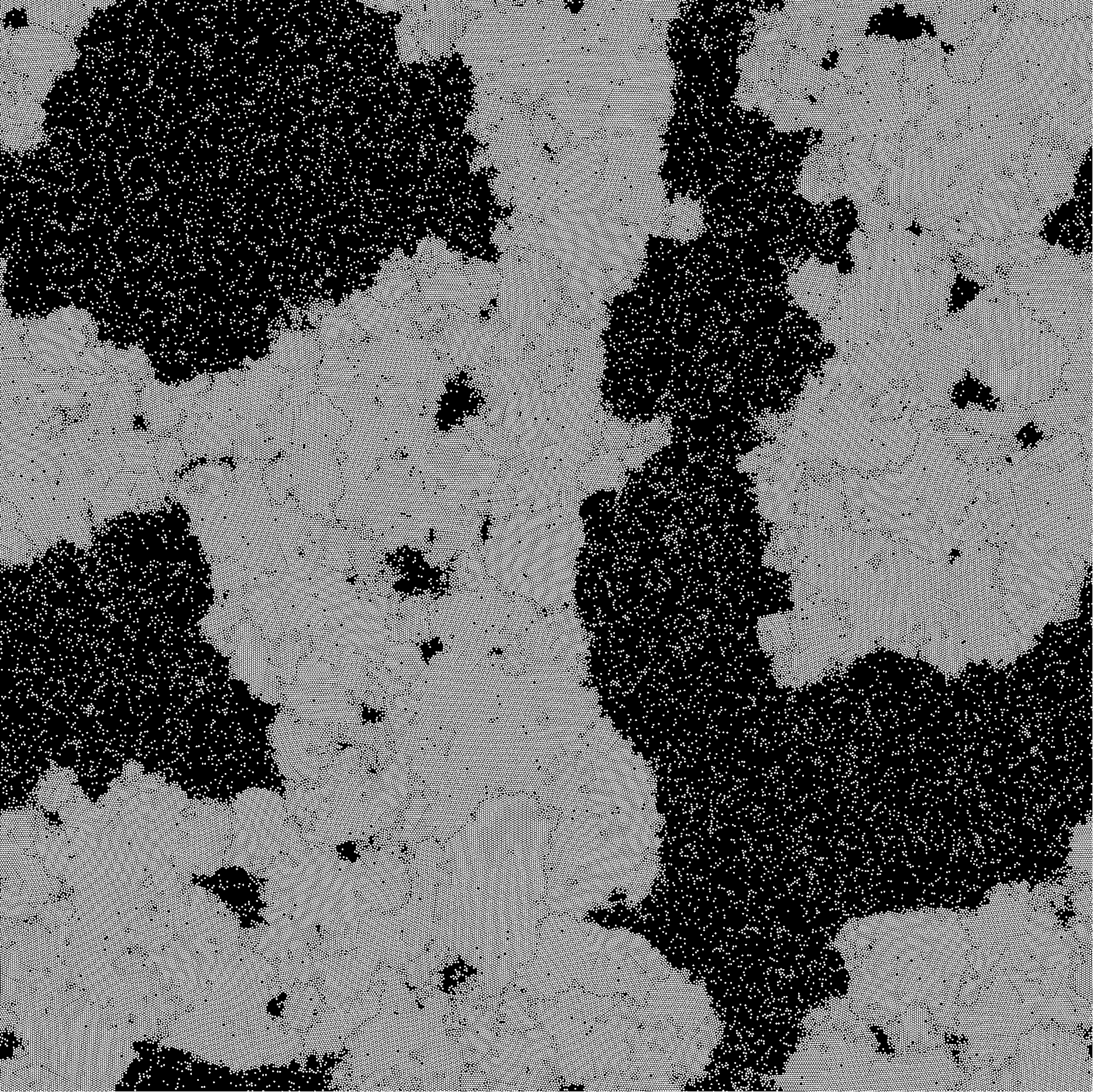}}
  \caption{Snapshots of a 2D ABP system with packing fraction $\phi=0.60$, alignment field strength $\tilde{B}=0$, and system size $512\sigma \times 512\sigma$ at various times after quenching from $\rm Pe=10$ to $\rm Pe=132$: $t\rm D_R=0.39$ (top left), $t\rm D_R=1.85$ (top right), $t\rm D_R=14.84$ (bottom left), and $t\rm D_R=337.36$ (bottom right). See also Video \ref{figSI:quench_movie_B0} in the SI.}
  \label{fig:quench_snapshots_B0}
\end{figure}
\begin{figure}
  \centering
  \subfloat{\includegraphics[width=.4\columnwidth]{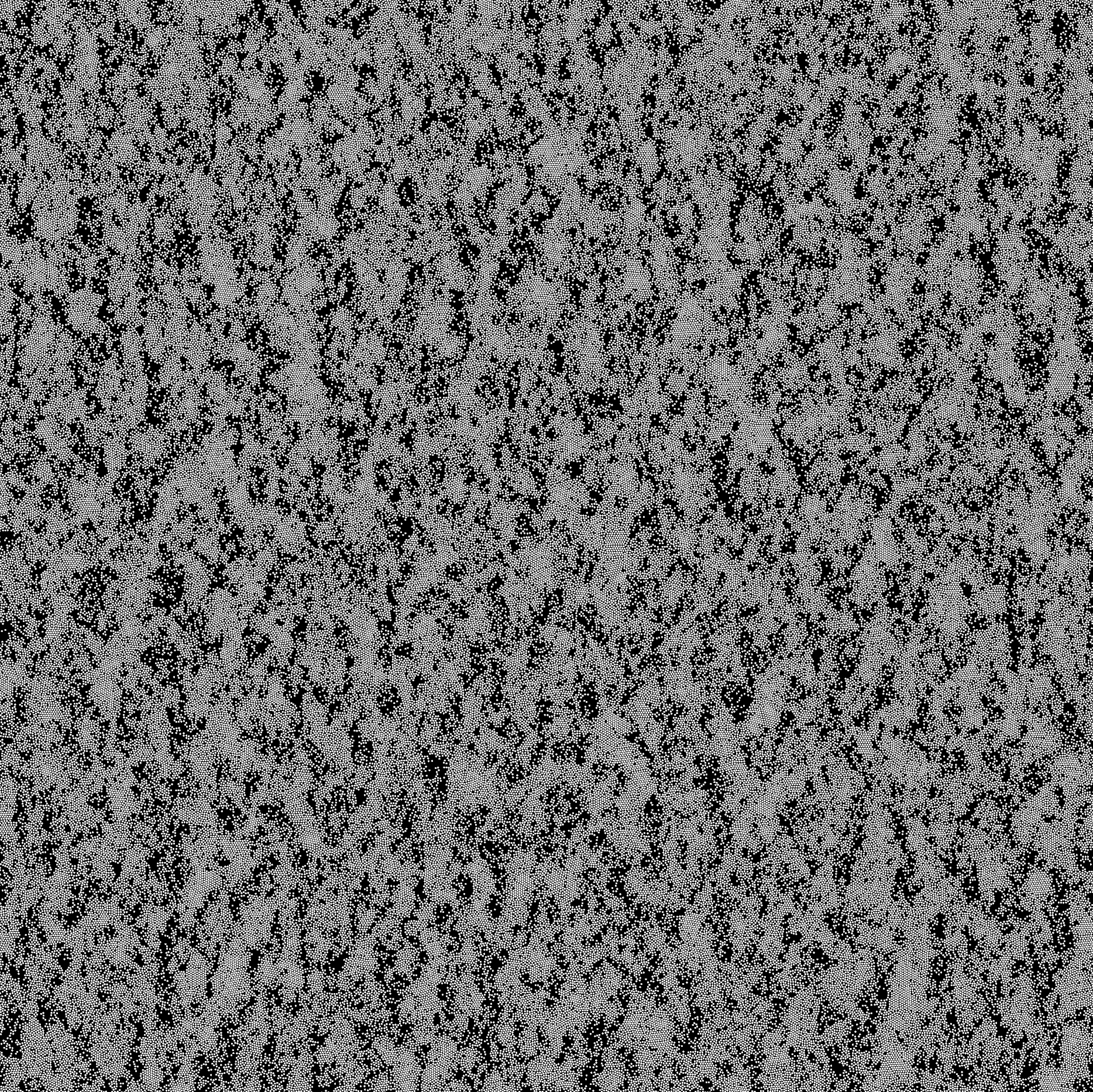}}\quad
  \subfloat{\includegraphics[width=.4\columnwidth]{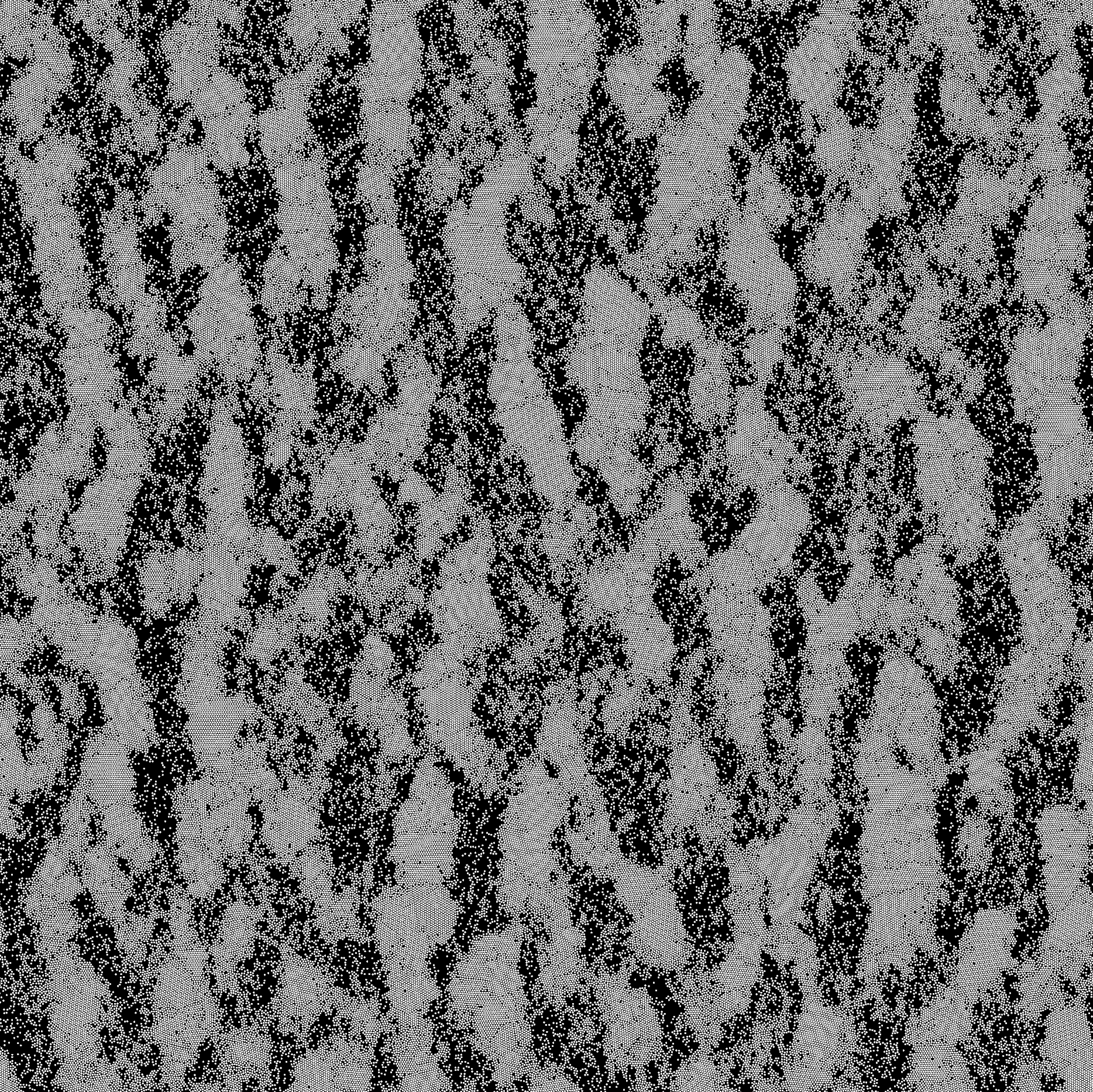}}\\
  \subfloat{\includegraphics[width=.4\columnwidth]{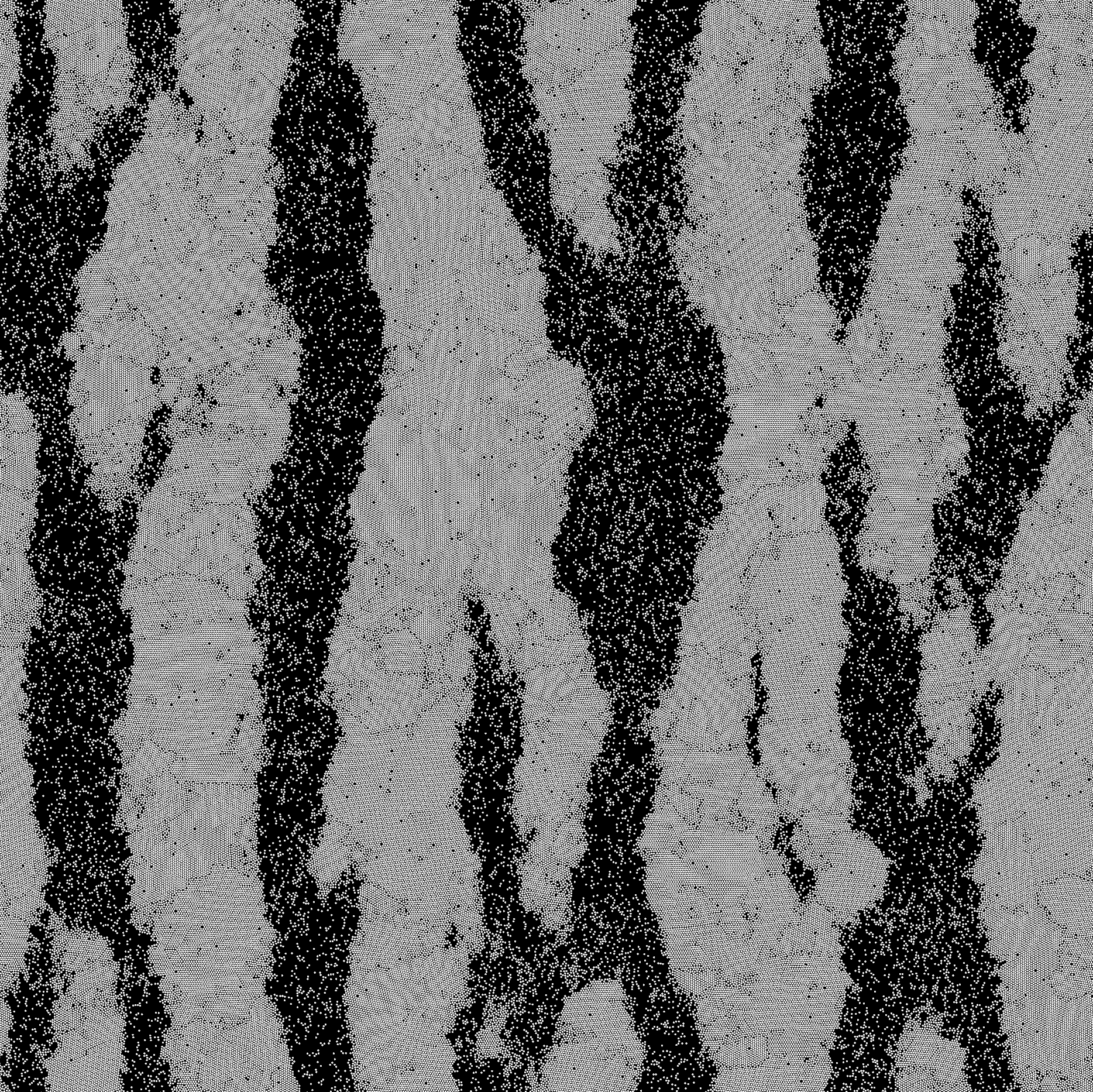}}\quad
  \subfloat{\includegraphics[width=.4\columnwidth]{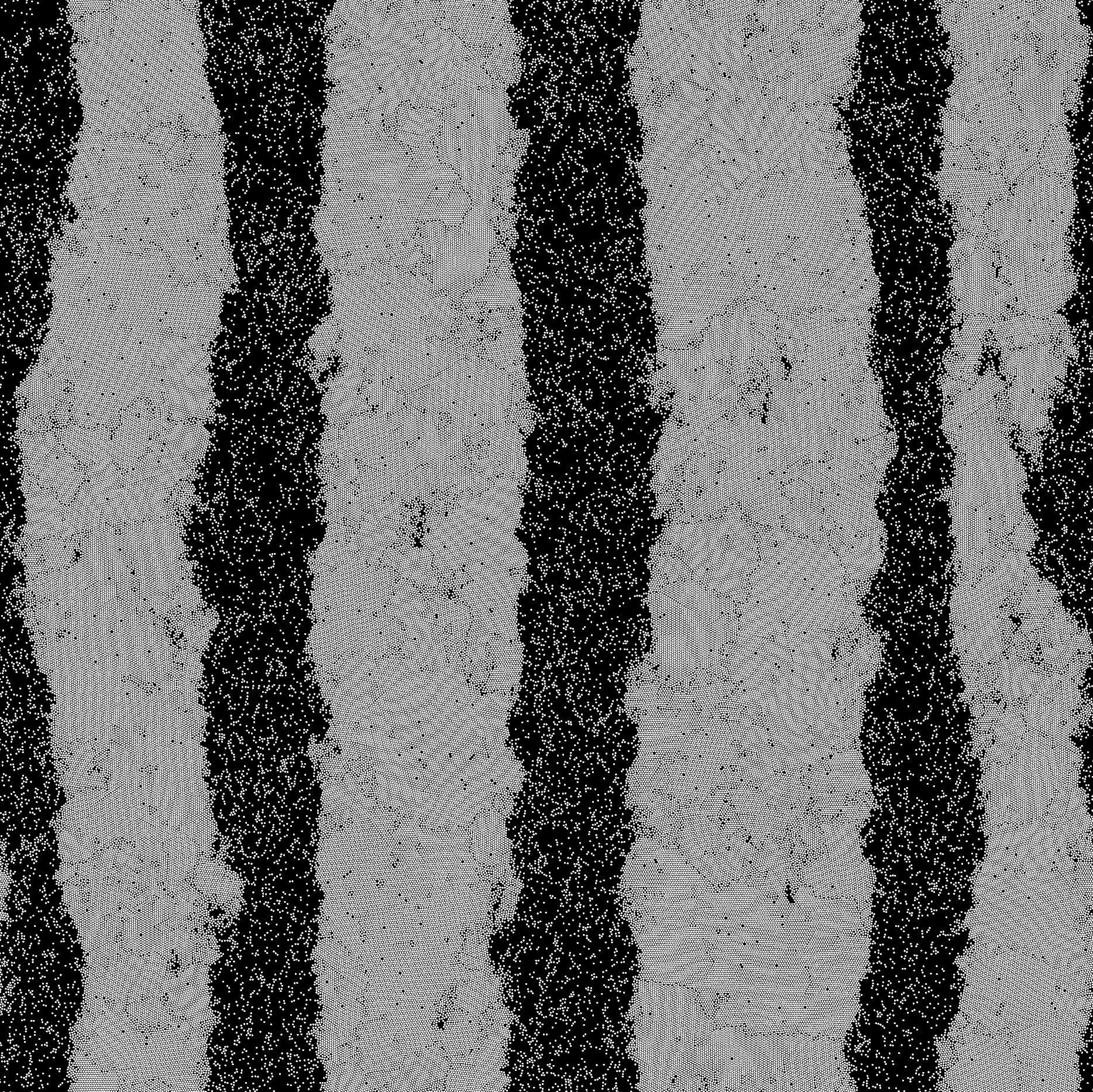}}
  \begin{tikzpicture}[overlay, remember picture]
    \draw[-stealth,line width = 3pt, red] (0.5, 2.5) -- (0.5, 5)  node[midway, above, sloped] {};
  \end{tikzpicture}
  \caption{Snapshots of a 2D ABP system with packing fraction $\phi=0.63$, alignment field strength $\tilde{B}=2$ (red arrow), and system size $512\sigma \times 512\sigma$ at various times after quenching from $\rm Pe=10$ to $\rm Pe=221$: $t\rm D_R=0.39$ (top left), $t\rm D_R=1.85$ (top right), $t\rm D_R=14.84$ (bottom left), and $t\rm D_R=337.36$ (bottom right). See also Video \ref{figSI:quench_movie_B2} in the SI.}
  \label{fig:quench_snapshots_B2}
\end{figure}
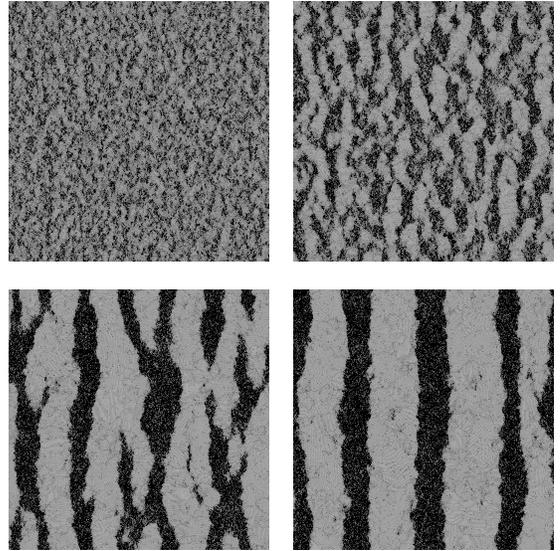

With the help of the power-law dependence of $\phi_{\rm liq}-\phi_{\rm gas}$ on $\tau$ in the range $0.3 < \tau < 0.7$, see Eq.~(\ref{eq:philmphig}), we find a critical exponent $\beta \approx 0.45$ for $\tilde{B}=0$, $2$, and $5$, see Fig.~\ref{fig:phase_diagram}(b). 
Thus, the critical exponent in the presence of a finite alignment field is similar to the critical exponent without.
Therefore, we hypothesize that the critical behavior found for the gas-fluid coexistence of active Brownian particles with and without alignment field belongs to the same universality class.
As reported earlier for active Brownian particles without alignment field \cite{siebert_critical_2018}, the critical exponent is much higher than $\beta_{\rm 2D\, Ising} = 0.125$ for a 2D Ising system and slightly lower than $\beta_{\rm MF} = 0.5$ for a mean-field model \cite{gallardo_analytical_2012}.

To determine the locations of the critical points more precisely, we calculate the cumulants \cite{siebert_critical_2018} 
\begin{equation}
    Q_{\ell_s} = \frac{\left< m^2 \right>_{\ell_s}^2}{\left< m^4 \right>_{\ell_s}}
    \label{eq:cumulant}
\end{equation}
following the procedure described in Ref.~\cite{midya_phase_2019}, see Fig.~\ref{fig:phase_diagram}(c).
Here, the second and the fourth moments of the order parameter are defined as
\begin{eqnarray}
   \left< m^n \right>_{\ell_{\rm s}} & = & \frac{1}{4} \left[ \sum_{i=1}^{2} \left( \left( \phi_{\rm gas, \ell_{\rm s}, i} - \phi_{\rm d} \right)^n + \left( \phi_{\rm liq, \ell_{\rm s}, i} - \phi_{\rm d} \right)^n \right) \right] \nonumber \\
\end{eqnarray}
with $n=2$ and $4$, respectively. The packing fractions $\phi_{\rm gas, \ell_{\rm s}, i}$ and $\phi_{\rm liq, \ell_{\rm s}, i}$ are  measured in sub-boxes of size $\ell_{\rm s} \times \ell_{\rm s}$ in the gas and the liquid phase, respectively; the index $i$ distinguishes the two stacked square boxes indicated in Fig.~\ref{fig:illustration}.
In the Supplementary Information (SI), we show that the slope of the cumulants $Q_{\ell_{\rm s}}$ with $\tau$ at the critical point increases as a power law with the sub-box size $\ell_{\rm s}$
\begin{equation}
    \left. d Q_{\ell_{\rm s}}/d\tau \right|_{\tau \simeq 0} \propto \ell_{\rm s}^{1/\nu}
\end{equation}
with $\nu \approx 1$, which is consistent with the value for the 2D Ising model.
Also, in this case, we find the same critical exponents for $\tilde{B}=2$ and $\tilde{B}=0$.

With increasing alignment field strength, the critical points shift to higher packing fractions $\phi_{\rm cr}$ and P\'eclet numbers $\rm Pe_{\rm cr}$. We fit the dependence of the coordinates of the critical points as
\begin{eqnarray}
\rm \phi_{cr} & = & 0.56 + 0.07 \, \tilde{B}^{1/3} - 0.016 \, \tilde{B}^{1/2} \label{eq:phicrVsB}\\
{\rm Pe_{cr}} & = & 32.23 + 19.09 \, \tilde{B} + 3.42 \, \tilde{B}^2 \label{eq:PecrVsB}
\end{eqnarray}
and collapse the phase boundaries for the two-phase coexistence region for various alignment field strengths, see Fig.~\ref{fig:phase_critical}. The good agreement of the normalized data supports the conclusion that the systems for all alignment field strengths belong to the same universality class.

\section{\label{sec:dynamics}Domain Coarsening Dynamics}
\begin{figure}
  \centering
  \begin{tabular}{@{}p{0.9\columnwidth}@{\quad}p{0.9\columnwidth}@{}}
    \subfigimg[width=\linewidth]{(a)}{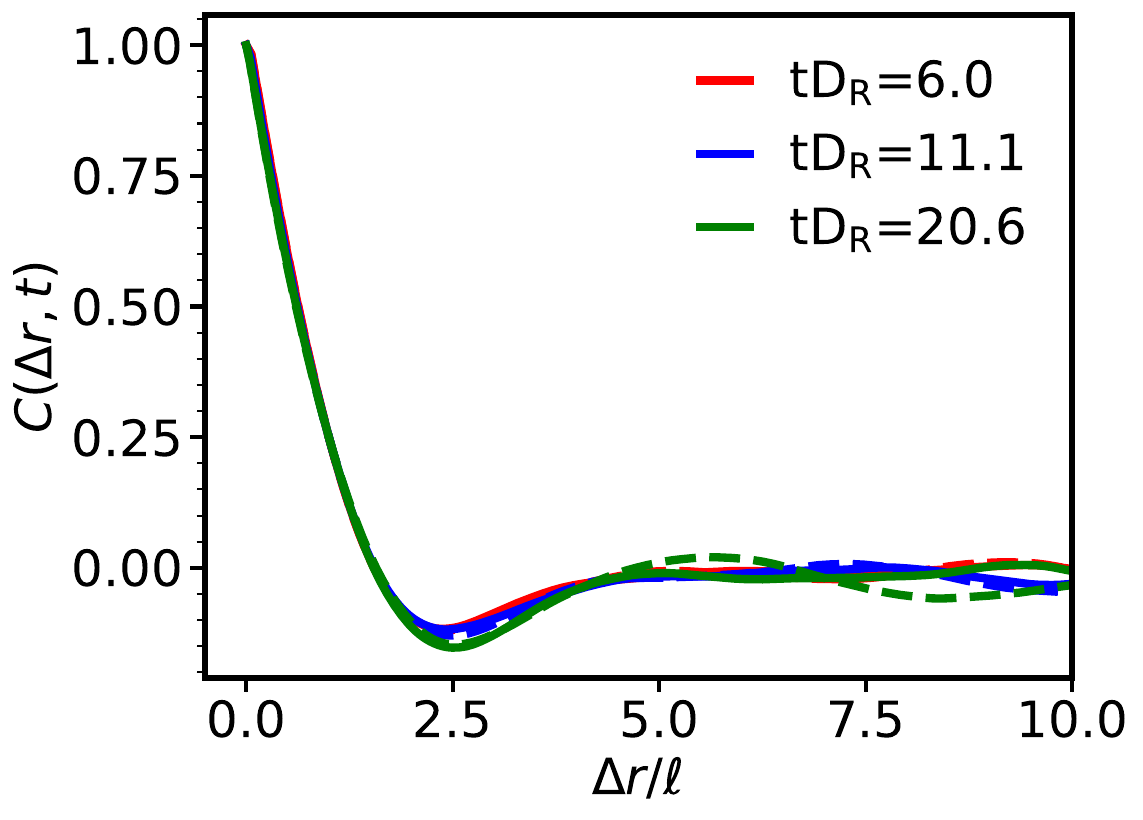} \quad \\
    \subfigimg[width=\linewidth]{(b)}{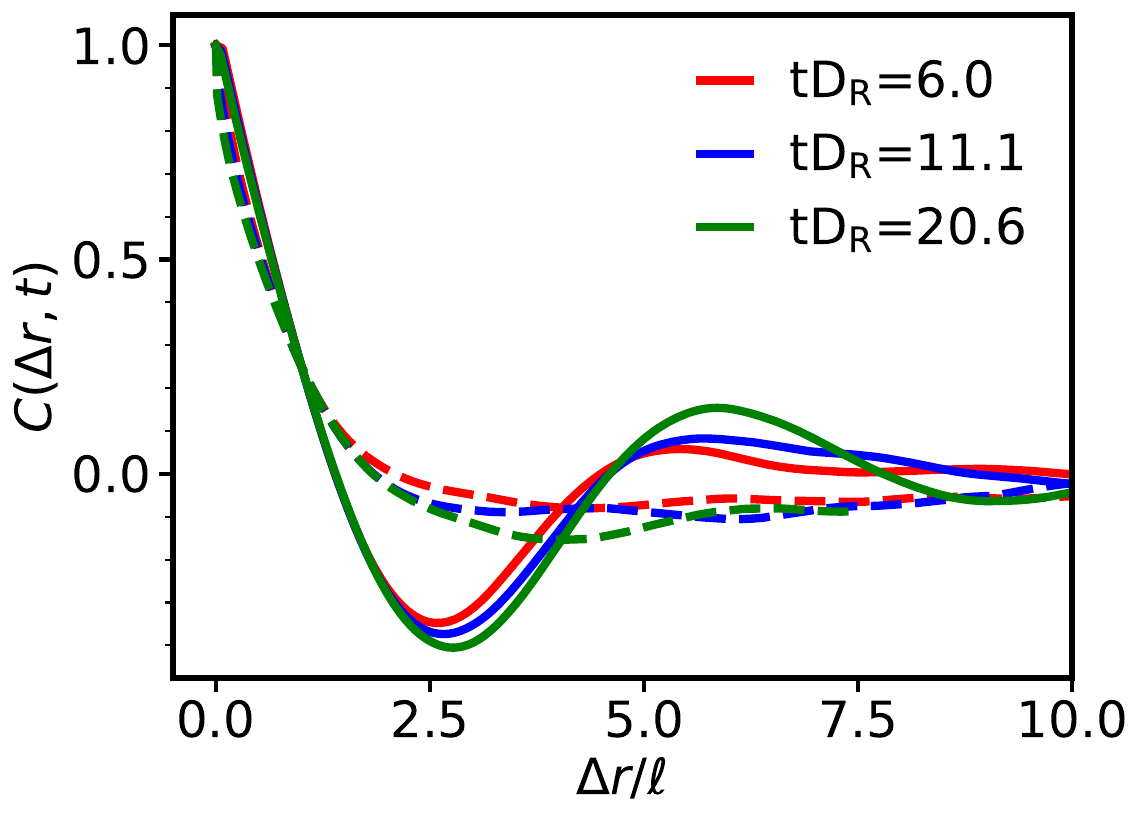} 
  \end{tabular}
  \caption{Two-point correlation functions $C({\bf r},t)$ at different times, scaled by the average domain length $\ell$, for (a) $\tilde{B}=0$ and (b) $\tilde{B}=2$. With an alignment field along the y-axis, the correlation function is close to one at different times. The solid line indicates the correlation function along the x-axis, and the dotted line along the y-axis.}
  \label{fig:dynamics_two_points_corr}
\end{figure}

\begin{figure}
    \centering
    \includegraphics[width=\columnwidth]{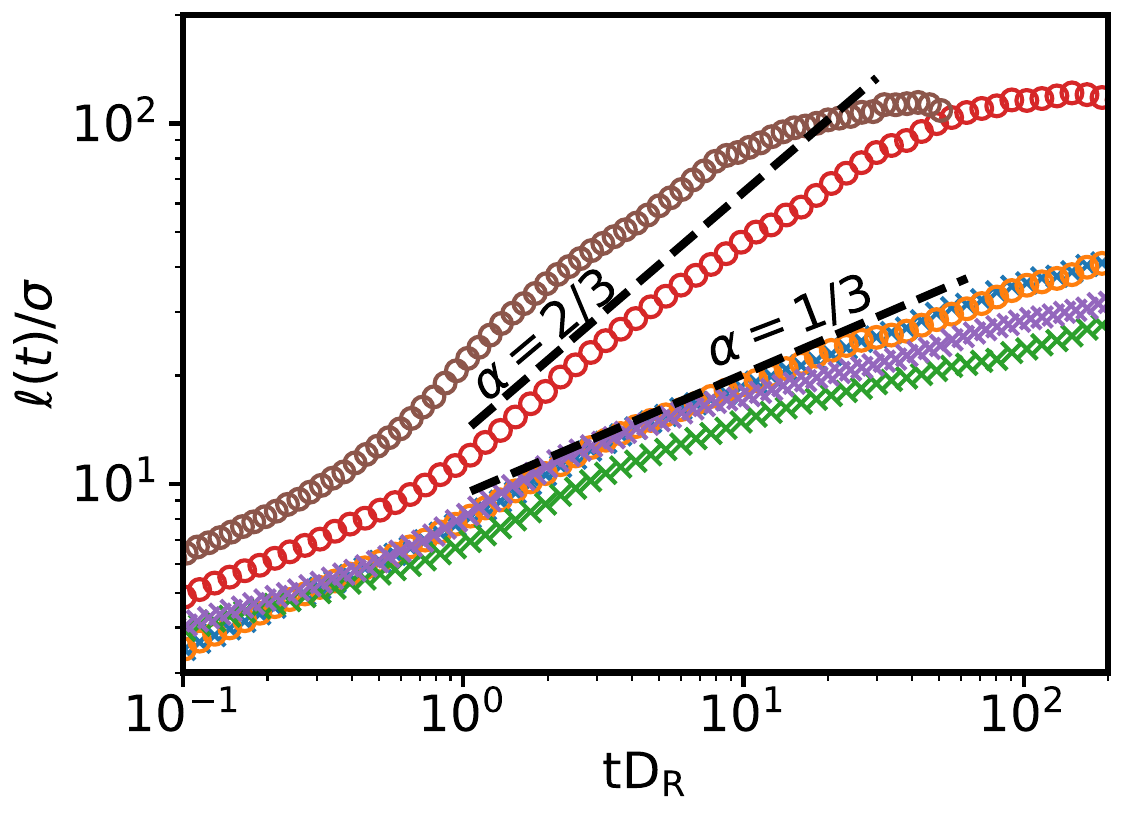}
    \caption{Domain length as a function of time for the quenched system, calculated from the two points correlation crossing at $C(r,t)=0.25$ in Fig.~\ref{fig:dynamics_two_points_corr}, for $\tilde{B}=0$ (blue, orange), $\tilde{B}=2$ (green, red), and $\tilde{B}=5$ (purple, brown). Circles symbols indicate the y-component, crosses the x-component.}
    \label{fig:compt_domain_length}
\end{figure}

Next, we investigate the effect of an alignment field on the kinetics of domain growth following quenches from outside ($\phi = \phi_{\rm cr}$, ${\rm Pe}=10$) to deep inside ($\phi = \phi_{\rm cr}$, ${\rm Pe}\simeq 3 \, {\rm Pe}_{\rm cr}$) the coexistence region, see Fig.~\ref{figSI:quench_illustration} in the SI.
Figure~\ref{fig:quench_snapshots_B0} shows the spinodal decomposition for $\tilde{B}=0$, Fig.~\ref{fig:quench_snapshots_B2} for $\tilde{B}=2$; videos can be found in the supplementary information. 
We characterize the dynamics using the equal time two-point spatial correlation function \cite{bray_theory_2002} 
\begin{eqnarray}\label{Eq-corr_fn}
C({\bf{r}},t) = \langle \psi({\bf{0}},t) \psi({\bf{r}}, t)\rangle - \langle \psi({\bf{0}},t)\rangle \langle \psi({\bf{r}},t)\rangle,
\end{eqnarray}
where $\psi({\bf{r}}, t)$ is the space- and time-dependent order parameter introduced in Sec.~\ref{sec:dyn_analysis_2p} of the SI. 
At early times, $tD_R \lesssim 1$, the domain patterns both for $\tilde{B}=2$ and  $\tilde{B}=0$ are isotropic in space and $C(x, y=0, t) \approx C(x=0, y, t)$, compare Figs.~\ref{fig:quench_snapshots_B0}(a), \ref{fig:quench_snapshots_B2}(a), and \ref{fig:dynamics_two_points_corr}. 
At later times, $tD_R > 1$, for $\tilde{B}=2$ first elongated clusters and eventually stripe patterns form in the direction of the alignment field; the correlations $C(x, y=0,t)$ transverse and $C(x=0, y,t)$ parallel to the direction of the alignment field direction differ. 
Interestingly, the average velocity of the particles in the direction of the alignment field in the gas phase is larger than in the liquid phase, i.e., $\langle v_{\rm y, gas}\rangle - \langle v_{\rm y, liq}\rangle > 0$, see Fig.~\ref{figSI:vgasmclus} in the SI. As a result, particles from the gas get deposited on the back of nucleated clusters.
The deviation of the simulation data in Fig.~\ref{fig:compt_domain_length} at late times from the power law is associated with the finite-size effects appearing much earlier in $\ell_y$ than in $\ell_x$. 

We characterize the sizes $\ell_y$ and $\ell_x$ of the domains parallel and perpendicular to the alignment field from the decays of $C(x=0,y,t)$ and $C(x,y=0,t)$ to $1/4$ of their initial values, respectively.
Whereas the domains perpendicular to the alignment field grow as $\ell_{\rm x} \propto t^\alpha$ with $\alpha=1/3$, similar to the field-free system, we find a much faster domain growth, $\ell_{\rm y} \propto t^{2/3}$, parallel to the field, see Fig.~\ref{fig:compt_domain_length}. 
The domain growth perpendicular to the field direction is dominated by the particle evaporation-condensation Lifshitz-Slyozov mechanism for Ostwald ripening, leading to $\ell_{\rm x} \propto t^{1/3}$ \cite{lifshitz_kinetics_1961}, which has been found previously for $\tilde{B}=0$  \cite{dittrich_growth_2023}. 
The faster growth $\ell_y \propto t^{2/3}$ hints at domain growth by merging of clusters, the exponent is similar to the scaling exponent $0.73$ for diffusion-limited cluster-cluster aggregation in $2$D \cite{caporusso_dynamics_2023}. 

Because the domain lengths show power-law growth $\ell \sim t^{\alpha}$, we expect the patterns to be statistically self-similar. Indeed, by dividing the correlation functions $C(x, y=0, t)$ and $C(x=0, y, t)$ by the corresponding average domain lengths $\ell_{\rm x}(t)$ and $\ell_{\rm y}(t)$, respectively, the correlation functions for fixed alignment field strength and direction collapse on master curves, see Fig.~\ref{fig:dynamics_two_points_corr}.
For $\tilde{B}=0$, the correlation functions decay, assume negative values for $2\lesssim \Delta r/\ell\lesssim 5 $, and remain approximately zero for $\Delta r/\ell \gtrsim 5$, see Fig.~\ref{fig:dynamics_two_points_corr}(a). For $\tilde{B}=2$, we find strong and long-reaching oscillations in $C(x, y=0,t)$, reflecting the stripe pattern being oriented along the $y$-direction parallel to the field, see Fig.~\ref{fig:dynamics_two_points_corr}(b). The correlation in the direction of the field, $C(x=0, y,t)$ decays without any overshoot to zero already for $\Delta r/\ell \gtrsim 3$.

\begin{figure}
  \centering
  \begin{tabular}{@{}p{0.9\columnwidth}@{\quad}p{0.9\columnwidth}@{}}
    \subfigimg[width=\linewidth]{(a)}{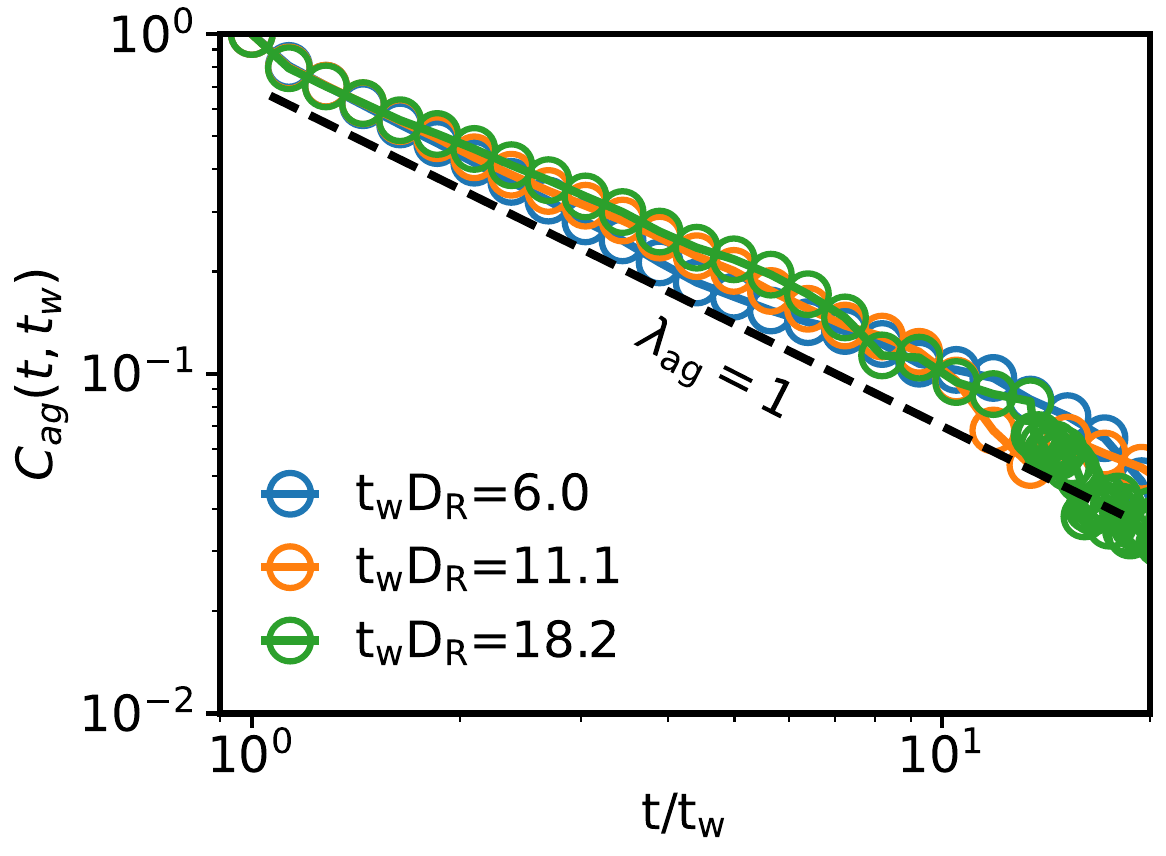} 
    \subfigimg[width=\linewidth]{\makebox[20pt][r]{(b)}}{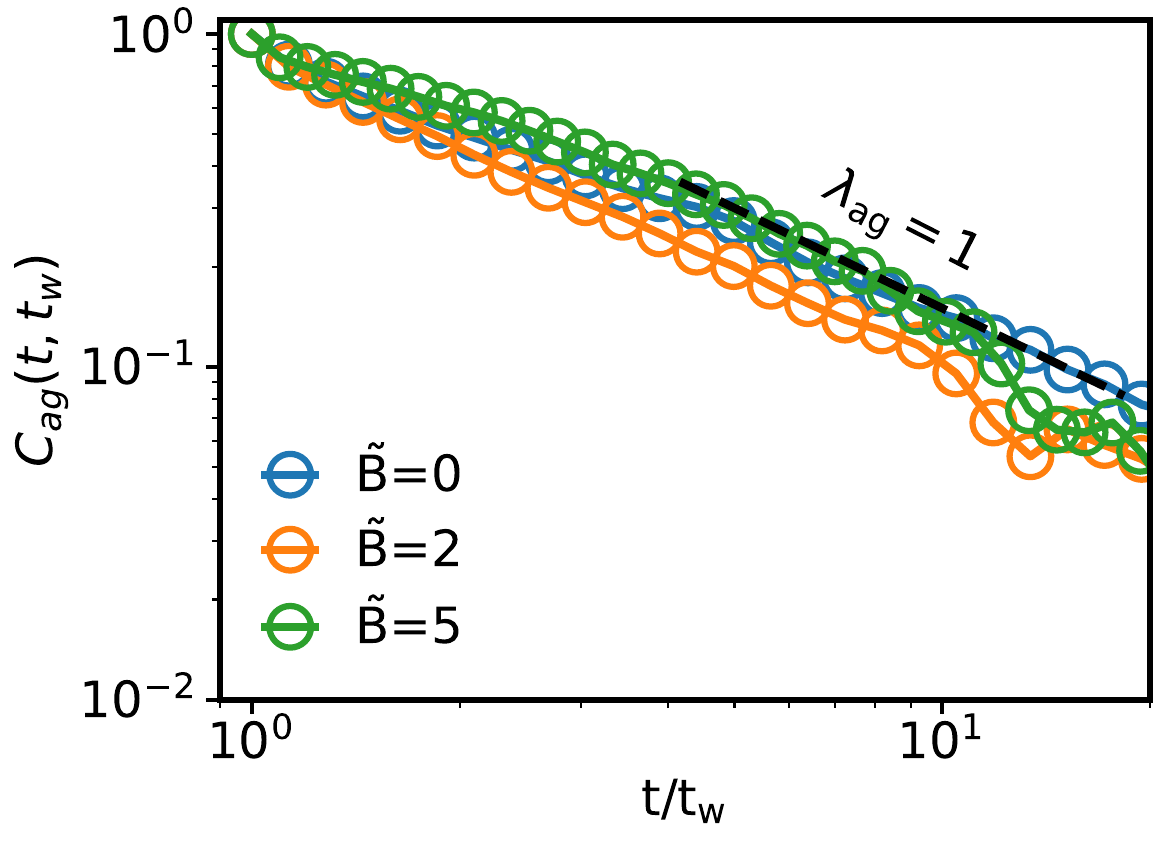} \quad   \\ 
  \end{tabular}
  \caption{Two-time correlation function $C_{\rm ag}(t, t_{\rm w})$ versus $t/t_w$ (a) for alignment-field strength $\tilde{B}=2$ and various waiting times and (b) for waiting time $t_{\rm w}D=11.1$ and various alignment-field strengths.}
  \label{fig:Auto_corr}
\end{figure}


The relaxation of the non-equilibrium ABP systems approaching steady states via domain coarsening can also be characterized using the two-time correlation function \cite{midya_aging_2014}
\begin{eqnarray}\label{Eq-auto_corr_fns}
 C_{\rm ag}(t,t_w) = \langle \psi({\bf{r}},t) \psi({\bf{r}}, t_w)\rangle - \langle \psi({\bf{r}},t)\rangle \langle \psi({\bf{r}},t_w)\rangle.
 \end{eqnarray}
Here, $t$ and $t_w$ (with $t>t_w$) are the observation time and the waiting time, respectively. When the system is in a steady state, $C_{\rm ag}(t,t_w)$ exhibits time-translational invariant properties, i.e., the data for $C_{\rm ag}(t,t_w)$ vs.\ ($t-t_w$) collapses for different choices of $t_w$. For out-of-equilibrium systems, the decay of $C_{\rm ag}(t,t_w)$ becomes slower with increasing waiting time $t_w$, thus, violates the above invariance. However, for the phase-separation kinetics of passive systems, Refs.~\cite{midya_dimensionality_2015} and \cite{midya_aging_2014} predicted a power-law scaling of $C_{\rm ag}(t,t_w)$ as a function of $t/t_w$ as 
\begin{eqnarray}\label{Eq-scaling-auto_corr_fns}
 C_{\rm ag}(t,t_w) \sim \left (\frac {t}{t_w} \right)^{-\lambda_{\rm ag}},
\end{eqnarray}
where the aging exponent $\lambda_{\rm ag}$ determines the relaxation rate of the non-equilibrium systems. 

For $\tilde{B}=0$, $C_{\rm ag}(t,t_w)$ as a function of $t/t_w$ for different waiting times $t_w$ collapses onto a single master curve with $\lambda_{\rm ag}=1$, see Fig.~\ref{figSI:Auto_corr} in the SI. The predicted value of $\lambda_{\rm ag}$ is consistent with the value reported in Ref.~\cite{midya_aging_2014}. 
The value of the aging exponent $\lambda_{\rm ag}$ is sensitive to various features, such as the conservation of order parameters and space dimensionality. For passive systems with conserved order-parameter dynamics, $\lambda_{\rm ag}$ satisfies the lower bound $\lambda_{\rm ag} \ge \alpha(\beta_s+d)/2$, where $d$ is the space dimensionality, and $\beta_s$ is associated with the small $k$ power-law behavior of the structure factor $S({\bf{k}},t)$, the Fourier transformation of $C({\bf{r}},t)$ \cite{yeung_bounds_1996}. For phase separating $2D$ ABP systems (with $\tilde{B}=0$), the small-$k$ power-law behavior of $S(\bf{k},t)$ is consistent with $\beta \simeq 3$ \cite{dittrich_growth_2023} which sets the lower bound $\lambda_{\rm ag} > 2.5\alpha \approx 0.8$ for $\alpha = 1/3$.

For $\tilde{B}=0$, the scaling of $C_{\rm ag}(t,t_w)$ as a function of $t/t_w$ is presented in Fig.~\ref{figSI:Auto_corr} in the SI. The master curve for the data from different values of $t_w$ follows a power-law decay with an exponent $\lambda_{\rm ag}=1$. The predicted value of $\lambda_{\rm ag}$ is consistent with the value reported in Ref.~\cite{midya_aging_2014}. 

For $\tilde{B}=2$, $C_{\rm ag}(t,t_w)$ as a function of $t/t_w$ shows strong periodic oscillations with decaying amplitudes for increasing times, see Fig.~\ref{figSI:Auto_correlation_osci} in the SI. The oscillations are associated with flow in the system along the field direction and the presence of periodic boundary conditions; see SI for more details.  
Therefore, we analyze the correlations after subtracting the average velocity of the particles along the field direction, see Fig.~\ref{fig:Auto_corr}(a).
The resulting master curve then decays with the same exponent $\lambda_{\rm ag}=1$ as in the case of $\tilde{B}=0$. 
We observe scaling of $C_{\rm ag}(t,t_w)$ for different alignment fields, and the master curve follows power-law decay with the same exponent $\lambda_{\rm ag}=1$, see Fig.~\ref{fig:Auto_corr}(b). 
This indicates that the relaxation dynamics of the $2$D ABP systems does not depend on the strength of the alignment field. 
As an alternative approach to subtracting the average particle velocity, we have analysed the temporal correlation function $C_{\rm dyn}(t,t_{\rm w})$ of the particle densities in the vicinity of selected particles, which we discuss in the SI. 
Interestingly, the master curves exhibit power-law decay $C_{\rm dyn}(t,t_{\rm w}) \sim (t/t_w)^{-\lambda^{\rm dyn}_{\rm ag}}$ with $\lambda^{\rm dyn}_{\rm ag}=2$, which indicates the presence of an alignment field for the direction of the propulsion velocity does not affect the relaxation dynamics of $2$D ABP systems.

\section{\label{sec:conclusions}Summary and Conclusions}

We have studied phase behavior and dynamics of domain growth in a system of $2$D active Brownian particles subject to a homogeneous external alignment field by systematically varying the strength of the alignment field. The binodals are estimated at different field strengths. The critical P\'eclet number and the critical packing fractions for gas-liquid coexistence increase with increasing field strength. The different binodals fall on a single master curve when the P\'eclet number is normalized with the critical P\'eclet number ${\rm Pe}_{\rm cr}$ and the packing fraction with the critical packing fraction $\phi_{\rm cr}$, which indicates that the order-parameter critical exponent $\beta$ is independent of the presence of an external field. Furthermore, we estimate  $\beta \approx 0.45$, which lies inbetween the values for the mean-field and $2$D Ising universality classes. Our prediction is consistent with the previously reported value for $2$D ABP systems without alignment field ($\tilde{B}=0$) \cite{siebert_critical_2018}.   

Furthermore, we have studied the dynamics of domain growth following quenches of the systems from outside to deep inside the coexistence region. The isotropicity and self-similarity of the evolution of percolating domain patterns are characterized via the 1D two-point correlation functions $C(x, y=0, t)$ and $C(x=0 , y, t)$. For $\tilde{B}=0$, in agreement with Ref.~\cite{dittrich_growth_2023}, we show that the domain growth occurs as $\ell_x \sim \ell_y \sim \ell \sim t^{1/3}$, following Lifshitz-Slyozov mechanism. For $\tilde{B}=2$, the formation of anisotropic domain patterns is observed, with stripe patterns parallel to the field direction. This leads to different power-laws for the domain growth parallel and perpendicular to the field. The domain growth of the domain size along the field direction follows the power-law $\ell_y \sim t^{2/3}$, whereas transverse to the field direction the growth follows the same power-law as for $\tilde{B}=0$, $\ell_x \sim t^{1/3}$. The decay of the autocorrelation function of the phase at a fixed point in space, $C_{\rm ag}(t,t_w) \sim (t/t_w)^{-\lambda_{\rm ag}}$ with exponent $\lambda_{\rm ag}=1$, characterizes the relaxation dynamics of the system. The decay exponent $\lambda_{\rm ag}$ does not depend on the alignment field strength, indicating that the  relaxation dynamics of $2{\rm D}$ ABP system does not change in the presence of the external field. 

In conclusion, ABPs that are subject to a homogeneous external alignment field move on average along the field direction, but--in a co-moving reference frame--can be characterized analogously to ABPs without an alignment field. Furthermore, the presence of the alignment field does not change the universality class of gas-liquid phase separation in 2D systems of ABPs, such that the critical points are characterized by the same critical exponents for different alignment field strengths and the coordinates of the critical points can be determined without the need for a cumulant analysis. 
However, the effective self-propulsion velocity and thus P\'eclet number of the particles in the co-moving reference frame
are reduced compared to alignment-field-free systems that 
have identical thermal noise and self-propulsion velocities in the lab reference frame. Therefore, an overall alignment of the 
direction motion does not only generate an overall particle flux, but also suppresses the formation of a dense liquid phase.

Directed motion and transport of active particles are ubiquitous in living and synthetic systems. The motion of
animal herds to new food sources, or the groups of soccer fans aiming for the stadium, are well-known examples.
For synthetic particles, alignment induced by external fields is the simplest way to control the overall motion of the particles in a system. In contrast to other techniques to generate an overall particle flux, such as structured channel boundaries \cite{ghosh2013self}, external fields also have the advantage that their strength and direction can be readily controlled and varied during an experiment.

\begin{acknowledgments}
JM thanks Peter Virnau (Mainz) and Suman Majumder (Amity University Noida, India) for discussions on critical behavior of active Brownian particles. SO acknowledges funding by the Bundesministerium für Bildung und Forschung (BMBF) within the Palestinian-German Science Bridge (PGSB). All authors acknowledge a computing time grant on the supercomputer JURECA at the Jülich Supercomputing Centre. We thank the Exploratory Research Space of RWTH Aachen University within the project ''Synthesizing Life-Like Material Systems” for fostering stimulating conceptual discussions.
\end{acknowledgments}

\appendix

\section{\label{sec:LAMMPS} Simulations parameters}

All simulations are preformed through molecular dynamics (MD) simulations using LAMMPS \cite{thompson_lammps_2022}. We simulate ABPs in rectangular simulation boxes with aspect ratio $1:3$ with various system sizes up to  $80\sigma\times240\sigma$ for cumulant analysis, and in square boxes of size $1024\sigma \times 1024\sigma$ for domain-coarsening dynamics. The particles are defined using the atom-style "hybrid dipole sphere" of LAMMPS, which adds a dipole moment to the particles whose orientation changes due to rotational diffusion and couples to an alignment field. 
The particle positions and dipole orientations are updated using the fix "Brownian-sphere". 
Throughout the simulation, we set $\rm T^*=Tk_B/\epsilon=1$, $\gamma_{\rm T}=\epsilon\sigma^2$, $\gamma_{\rm r}=0.40839\epsilon\sigma^2$, and use independent Gaussian-distributed noises $\boldsymbol{\xi}_T$ and $\boldsymbol{\xi}_R$ with unit variance. 

Initially, the particles are randomly distributed using a new random seed for every run and restart of a simulation. For new simulations, the "minimize" function is used to ensure that the particles are not overlapping. 
We start with a run of $10^{6}$ time steps to obtain a steady state. The integration time step is $10^{-6}s$. 
For a full production run, we then simulate for $800*10^{6}$ time steps. All analysis of LAMMPS dump files have used OVITO \cite{stukowski_visualization_2009} and Python.

\bibliography{ActiveParticlesInChannels}

\end{document}